\documentclass{appolb}

\usepackage{epsfig}
\usepackage{graphicx}
\begin{document}

\title{Restoration of axial symmetry  and its possible relation  with restoration 
of chiral symmetry and deconfinement  at finite temperature 
\thanks{Presented at eQCD2013}%
}

\author{M. C. Ruivo, P. Costa, C. A. de Sousa
\address{Centro de F\'{\i}sica Computacional, Departamento de F\'{\i}sica,
Universidade de Coimbra, P-3004 - 516 Coimbra, Portugal}
}
\date{\today}

\maketitle

%%%%%%%%%%%%%%%%%%%%%%%%%%%%%%%%%%%%%%%%%%%%
%% ABSTRACT
%%%%%%%%%%%%%%%%%%%%%%%%%%%%%%%%%%%%%%%%%%%%

\begin{abstract}

The phase transitions characterized by deconfinement and restoration of chiral symmetry as well as 
the restoration of  axial symmetry, at finite temperature,  are investigated in the framework of  
SU(2) Polyakov-Nambu-Jona-Lasinio (PNJL) models with the U$_A$(1) anomaly.   
The thermodynamics  of the phase transitions, the topological susceptibility, the meson spectrum, 
and, in particular, the convergence of axial and chiral partners are  analyzed, in the framework 
of the ordinary PNJL model and in  its extension, the entangled Polyakov-Nambu-Jona-Lasinio (EPNJL) model. 
The latter incorporates  entanglement between restoration of chiral symmetry and deconfinement.

\end{abstract}

%%%%%%%%%%%%%%%%%%%%%%%%%%%%%%%%%%%%%%%%%%%%
%% MAINMATTER
%%%%%%%%%%%%%%%%%%%%%%%%%%%%%%%%%%%%%%%%%%%%
\vskip0.25cm

An important  field of research nowadays, both in  experimental and theoretical physics, 
is the study of matter under  extreme conditions of temperatures and/or  densities.
Quantum chromodynamics (QCD) predicts that,  in this limit, matter becomes a plasma of 
deconfined quarks and gluons. In spite of the success of many aspects of the strong
interaction physics, there are important features to clarify, and, in this context,  
the use of QCD inspired effective models has proven to be a useful approach.

The Polyakov-Nambu-Jona-Lasinio model is an effective model which respects important 
symmetries of the QCD action. It contains quarks as fundamental degrees of freedom allowing 
for a self-consistent description of chiral symmetry breaking and restoration. 
In addition,  the coupling to the Polyakov loop allows to describe the (statistical) 
confinement/deconfinement phase transition by taking into account a static gluonic field in which 
quarks propagate \cite{Meisinger:1996PLB,Ratti:2005PRD}. 
An important query on QCD thermodynamics is the proximity or coincidence of the chiral 
and deconfinement phase transitions. 
In this concern, lattice QCD results have been a matter of debate:  the two phase transitions 
are reported to occur at distinct temperatures  for $N_f=2+1$ flavors, but at the same temperature, 
$T_c=174(3)(6)$ MeV,  for $N_f=2$ flavors \cite{lattice_Tc3F}.  
This effect could be the result of strong correlations (entanglement) between the quark condensate 
and the Polyakov field, $\Phi$, an effect that is incorporated in the 
entangled Polyakov--Nambu--Jona-Lasinio model \cite{Sakai1}.

The restoration of the  U$_A$(1) symmetry  is also a longstanding question and phenomenological 
consequences for the nature of the phase transition are expected to occur, depending on  the degree 
of anomaly present at the critical temperature.
Moreover, the topological susceptibility, the meson axial chiral partners and  the $\eta'$ mass 
can exhibit  signs of the restoration of the U$_A$(1) symmetry. 
Lattice calculations with three flavours have found evidence for the decrease of the topological 
susceptibility with temperature  and for convergence of the  meson correlators of chiral and
axial partners   when both symmetries are restored. 
The return of the 9th ``prodigal'' Goldstone boson  has been a matter of debate and recent 
experimental results  are compatible with a decrease of  about 200 MeV for the in medium mass of 
the $\eta'$ meson (see \cite{Ruivo122} and references therein).

QCD inspired models  have  been used to study the restoration of chiral and axial symmetries, 
in particular the NJL model  and its extended version, the PNJL model \cite{Costa}.  
In previous investigations, both in   NJL model and PNJL SU(3) models  \cite{Regulariz}, 
it has been found that  observables related with the U$_A$(1) symmetry breaking vanish
as a natural consequence of the effective  restoration of chiral symmetry, provided 
a specific regularization  is used \cite{Regulariz}. 
However, the situation is different in the framework of the  SU(2)  models
(see \cite{Ruivo122,Santos}) where an additional  mechanism of instantons suppression is needed.

%%%%%%%%%%%%%%%%%%%%%%%%%%%%%%%%%%%%%%%%%%%%%%%%%%%%%%%%%%%%%%%%%%%%%%%
%%%%%%%%%%%%%%%%%%%%%%%%%%%%%%%%%%%%%%%%%%%%%%%%%%%%%%%%%%%%%%%%%%%%%%%
%%%%%%%%%%%%%%%%%%%%%%%%%%%%%%%%%%%%%%%%%%%%%%%%%%%%%%%%%%%%%%%%%%%%%%%
%%%%%%%%%%%%%%%%%%%%%%%%%%%%%%%%%%%%%%%%%%%%%%%%%%%%%%%%%%%%%%%%%%%%%%%

\vskip0.40cm
The SU(2) PNJL Lagrangian  with a 't Hooft interaction simulating the U$_A$(1) anomaly 
\cite{Ratti:2005PRD,Santos} is given by :

%%%%%%
\begin{eqnarray}
\mathcal{L_{PNJL}} &=& \bar{q}(\,i\, {\gamma}^{\mu}\,D_\mu\,-\,\hat m)q + \mathcal{L}_1 +
\mathcal{L}_2\,-\,\mathcal{U}\left(\Phi[A],\bar\Phi[A];T\right),
\label{e1}
\end{eqnarray}
%%%%%%
with two different interacting parts
%%%%%%
\begin{eqnarray}
\mathcal{L}_1 &=& g_1 \Big{[}(\overline{q} q)^2 + (\bar{q} i \gamma _5 \vec{\tau} q)^2 +
(\bar{q} \vec{\tau} q)^2 + (\bar{q} i \gamma _5 q)^2 \Big{],}
\label{e1_la}
\end{eqnarray}
%%%%%%
%%%%%%
\begin{eqnarray}
\mathcal{L}_2 &=& g_2 \Big{[}(\bar{q} q)^2 + (\bar{q} i \gamma _5 \vec{\tau} q)^2 -
(\bar{q} \vec{\tau} q)^2 - (\bar{q} i \gamma _5 q)^2 \Big{],}
\label{e1_lb}
\end{eqnarray}
%%%%%%
where  $q = (u,d)$ are the Dirac and color fields,
with two flavors, $N_f=2$, and three colors, $N_c=3$, the coupling coefficients   $g_1$ and
$g_2$ have  dimension $energy^{-2}$, and  $\hat{m}=\mbox{diag}(m_u,m_d)$ is the current
quark mass matrix, here being assumed that $m_u=m_d=m$. Both terms $\mathcal{L}_1$ and $\mathcal{L}_2$  
are invariant upon SU(2)$_{L}\otimes$SU(2)${_R}\otimes$U(1) type transformations, but  $\mathcal{L}_2$, 
that  can be  written as a determinant (see \cite{Santos}), is not invariant upon U$_A$(1) transformations. 

The Lagrangian density  (\ref{e1}) can be rewritten as
%%%%%%
\begin{eqnarray}
{\mathcal L_{PNJL}\,}&=& \bar q\,(\,i\, {\gamma}^{\mu}\,D_\mu\,-\,\hat m)\,q \nonumber\\ 
&+& {G_s}\, [(\bar{q} q)^2 + (\bar{q} i \gamma _5 \vec{\tau} q)^2 ] +
{G_a}\,[(\bar{q} \vec{\tau} q)^2 + (\bar{q} i \gamma _5 q)^2 ] \nonumber\\
&-& \mathcal{U}\left(\Phi[A],\bar\Phi[A];T\right), \label{eq:lag}
\end{eqnarray}
%%%%%%
where  $G_s=g_1+g_2$ and $G_a=g_1-g_2$ enter, respectively, in the propagators of the meson 
chiral partners  $(\pi,\,\sigma)$  and  $(\eta,\,a_0)$. 
We can redefine the coupling constants such as the  set ($g_1, g_2$) or ($G_s, G_a$) will
be replaced by ($ G,\, c$) in the following parametrization:
%%%%%%
\begin{equation}
G_s= g_1 + g_2 = G, \,\,\,\,\,\,\,\,\, G_a= g_1 - g_2 = G\, (1\,-\, 2\,c),
\label{redef}
\end{equation}
%%%%%%
where  $c \,\in \,\{0,1\} $  is a parameter that now specifies the degree  of  U$_A$(1) symmetry breaking. 
Notice that $g_1=  G\,(1- c)$ is associated with chiral symmetry effects, while $g_2= c \,G$ is 
the anomaly coefficient and is here $25\%$ of $g_1$. The  parameters in the NJL sector of the model are  
$m=6$ Mev, $\Lambda=590$ Mev, $G\Lambda^2=2.435$,  $c=0.2$, and were fixed by fitting lattice or experimental 
values for the pion mass, its decay constant, the quark condensate and the topological susceptibility, $\chi$. 
The masses of the $\sigma$, $\eta$ and $a_0$  mesons come as outputs (for detail see \cite{Ruivo122,Santos} 
and references therein).

The quarks are coupled to the gauge sector {\it via} the covariant derivative $D^{\mu}=\partial^\mu-i A^\mu$, 
where $A^\mu(x) = g_{Strong} {\cal A}^\mu_a(x)\frac{\lambda_a}{2}$ and ${\cal A}^\mu_a$ is the SU$_c(3)$ 
gauge field and $\lambda_a$ are the Gell--Mann matrices.  
The Polyakov loop $\Phi$  is the trace of the Polyakov line defined by:
$ \Phi = \frac 1 {N_c} {\langle\langle \mathcal{P}\exp i\int_{0}^{\beta}d\tau\,
A_4\left(\vec{x},\tau\right)\ \rangle\rangle}_\beta$.

The effective potential for the Polyakov loop is:
%%%%%%
\begin{equation}
    \frac{\mathcal{U}\left(\Phi,\bar\Phi;T\right)}{T^4}
    =-\frac{a\left(T\right)}{2}\bar\Phi \Phi +
    b(T)\mbox{ln}[1-6\bar\Phi \Phi+4(\bar\Phi^3+ \Phi^3)-3(\bar\Phi \Phi)^2].
    \label{Ueff}
\end{equation}
%%%%%%

The parameters of the effective potential $\mathcal{U}$ have been fixed in order to reproduce the lattice
data for the expectation value of the Polyakov loop and QCD thermodynamics in the pure
gauge sector  and are:  $a_0=3.51, a_1=-2.47, a_2= 15.2, b_3=-1.75$; the parameter $T_0$ is  the critical 
temperature for the deconfinement phase transition within a pure gauge approach and it was fixed to $270$ MeV.

%%%%%%%%%%%%
%EPNJL MODEL
%%%%%%%%%%%%
The PNJL model with the  original $T_0 = 270$ MeV yields to a small difference between chiral and 
deconfinement transition temperatures.
This value is however significant when we rescale $T_0$ to 210 MeV, derived by RG considerations   
so as to reproduce the lattice QCD result, $T_d=177$ MeV. 
Consequently, the PNJL result is not consistent with lattice QCD data for the transition temperatures 
and  entanglement between chiral and deconfinement transitions is weak in this model. 
The EPNJL model  \cite{Sakai1} incorporates entanglement by endowing the four-quark coupling vertex 
$G$ with a dependence on the Polyakov field, $\Phi$, in the form:
%%%%%%%%
\begin{equation}
G(\Phi) = G\left[1\,-\,\alpha_1\,\Phi \bar\Phi\,-\,\alpha_2\,( \Phi^3\,+\,\bar\Phi^3) \right],
\label{gphi}
\end{equation}
%%%%%%%%
which respects chiral, $P$, $C$ and the extended center symmetries. 
The parameters $\alpha_1=\alpha_2=0.2$ and $T_0 = 170$ MeV, taken from  reference \cite{Sakai1}, 
were fixed to reproduce the available lattice QCD data.

As it can be seen in Table 1, in the PNJL model $T_{\chi}$ and $T_d$ never coincide but are closer 
for higher values of $T_0$, so it is adequate to  choose $T_0 \simeq 270$ MeV.
In the EPNJL model, where $\Delta=(T_{\chi}-T_d)/T_{\chi}=0$ by construction, we have more freedom 
to fix $T_0$ and a lower value is convenient $(T_0\simeq 170\mbox{ MeV})$ since it allows to reproduce 
lattice results for the critical temperature.
The results are presented for scenario B, for reasons that will be explained latter.
%%%%%%%%%%%%%%%%%%%%%%%%%%%%%%%%%%%%%%%%%%%%%%%%%%%%%%%%%%%%%%%%%%%%%%%
\begin{table}[t]
    \begin{center}
        \begin{tabular}{||c||c|c|c|c|c||}
          \hline\hline
            Scenario B 	& $T_0$ & $T_{\chi}$	& $T_d$	& $\Delta$ 	& $T_{eff}$ \\
            						& [MeV] & [MeV] 			& [MeV]	& 	--			& [MeV] 		\\
          	\hline
            \hline
 %           PNJL   & 200 & 216 & 170 & 22\% & ... \\
             PNJL   & 210 & 215 & 177 & 18\% & $\sim$ 250 \\
             $ $   & 270 & 237 & 219 &  8\% & $\sim$ 300\\
            \hline
            EPNJL  & 170 & 173 & 173 & -- & $\sim$ 200\\
             $  $  & 270 & 223 & 223 & -- & $\sim$ 300\\
       \hline\hline
        \end{tabular}
         \caption{
          Characteristic temperatures in PNJL and EPNJL model for different values of
          $T_0$ $\left(\Delta=(T_{\chi}-T_d)/T_{\chi}\right)$. $T_{eff}$ is the temperature 
          at which the effective restoration of both symmetries is achieved.}
    \end{center}
\label{table:Temp}
\end{table}
%%%%%%%%%%%%%%%%%%%%%%%%%%%%%%%%%%%%%%%%%%%%%%%%%%%%%%%%%%%%%%%%%%%%%%%

Notice that  at finite $T$, due to the presence of the partition functions, the finite cutoff, 
$\Lambda$, used in the vacuum  to regularize integrals,  is not necessary and we  make 
$\Lambda \longrightarrow \infty$. 
This procedure  ensures that the pressure goes  to the Stefan--Boltzmann limit and  
leads to a better description of several thermodynamic quantities. It  has the disadvantage of 
leading to a sharp decrease of the quark condensates that do not vanish assimptoticaly and 
change sign unless a mechanism is imposed by hand to avoid this (for details see \cite{Regulariz}).
In SU(3) NJL and PNJL models the effective restoration of  axial symmetry occurs as consequence 
of the full restoration of chiral symmetry. As a matter of fact, since in those models the 't Hooft 
interaction is a six-quark interaction, that in the usual approach is reduced to a four-quark 
interaction by means of a Wick contraction of a quark-antiquark pair, we have  a kind of 
``effective'' 't Hooft coupling coefficient, $\tilde{G}_D=G_D \,<\bar qq>$,  and, therefore, 
the anomaly effects disappear when the quark condensate anihilates.  This is  not so in the SU(2) 
models where the 't Hooft interaction is a four-quark vertex.
In SU(2) models (even in the EPNJL model where the coupling vertices have a dependence of 
temperature through the Polyakov field, $\Phi$)  a fraction of the anomaly remains in the 
chiral restored phase because, although the topological susceptibility vanishes, the meson 
axial partners do not converge. In order to get  this convergence  it is  enough that only  
$g_2 $ acquires the adequate dependence on temperature and  $g_1 $ is kept constant 
(scenario {A}, see \cite{Ruivo122}). 
%%%%%%%%%%%%%%%%%%%%%%%%%%%%%%%%%%%%
\begin{figure}[t]
\begin{center}
  \begin{tabular}{cc}
       \hspace*{-0.5cm}\epsfig{file=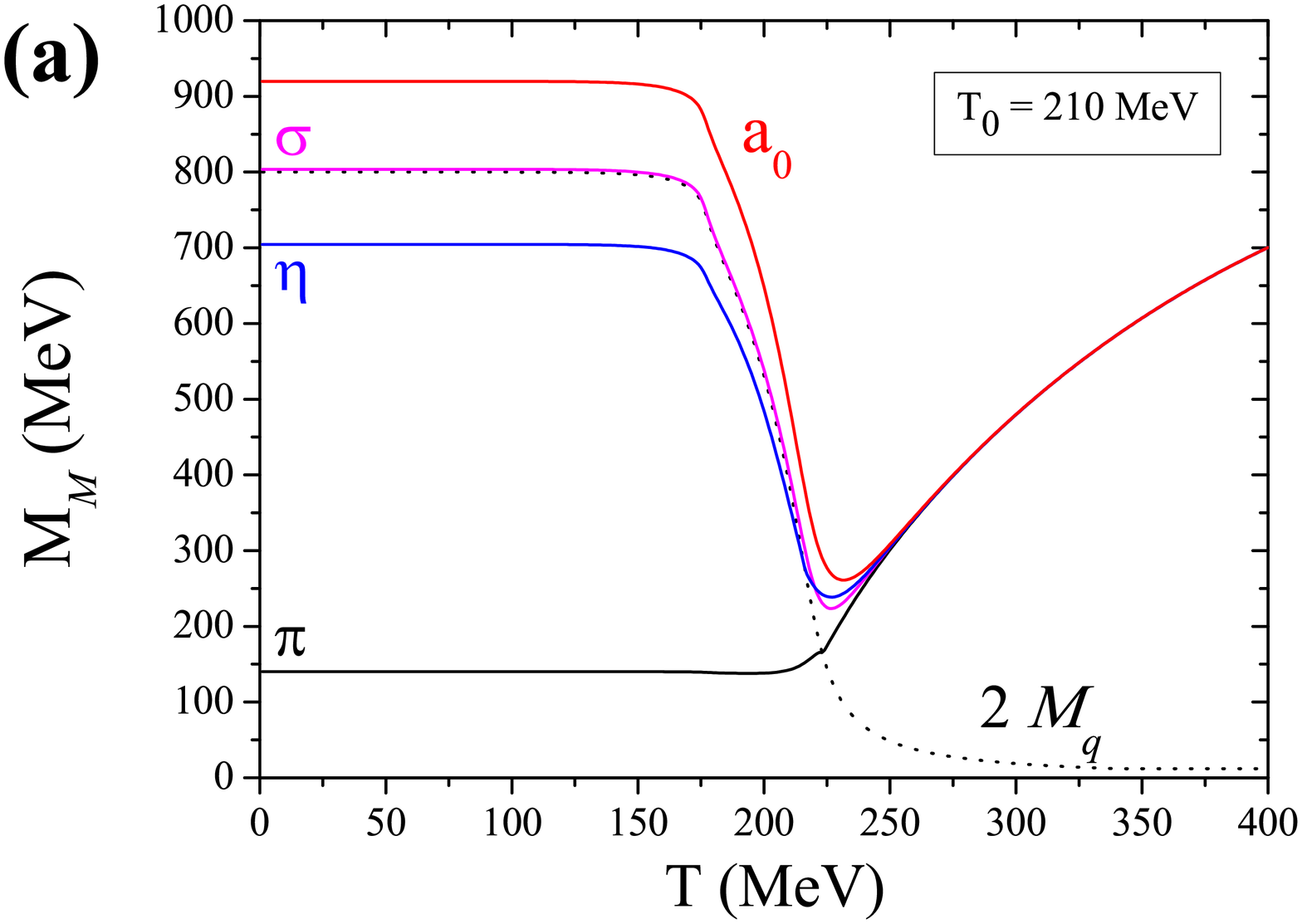,width=5.5cm,height=4.5cm}&
       \hspace*{-0.5cm}\epsfig{file=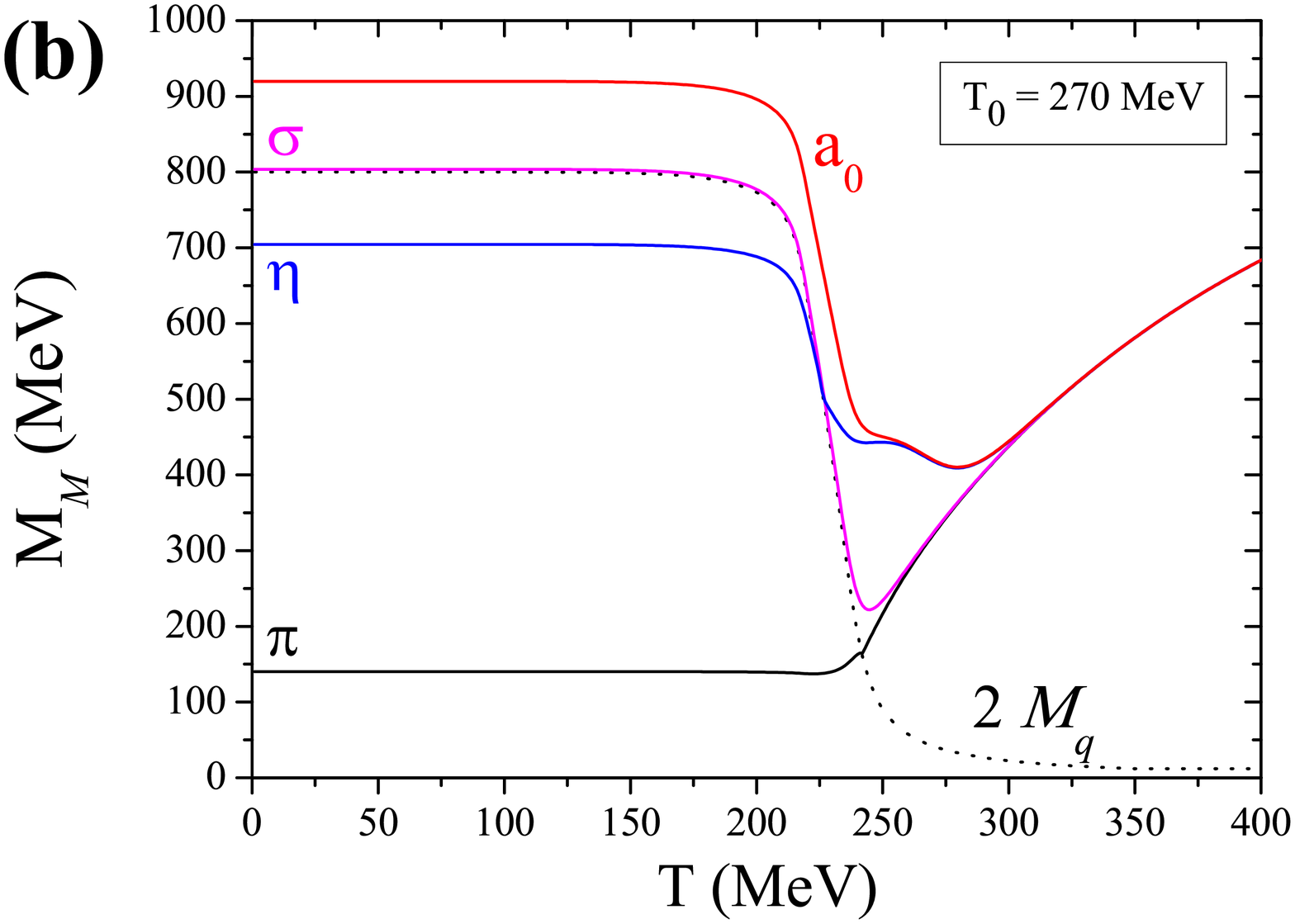,width=5.5cm,height=4.5cm}\\
       \hspace*{-0.5cm}\epsfig{file=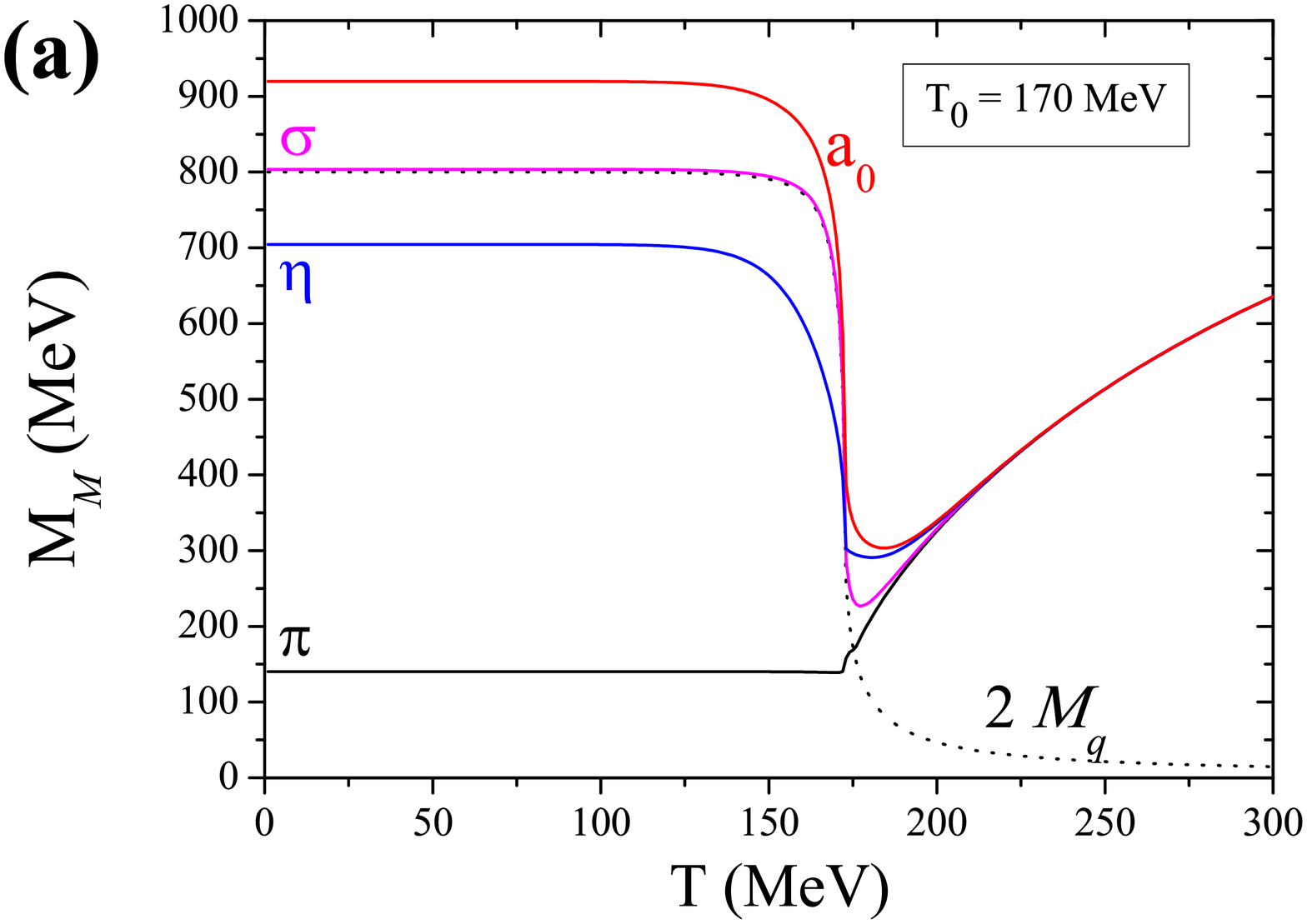,width=5.5cm,height=4.5cm}&
       \hspace*{-0.5cm}\epsfig{file=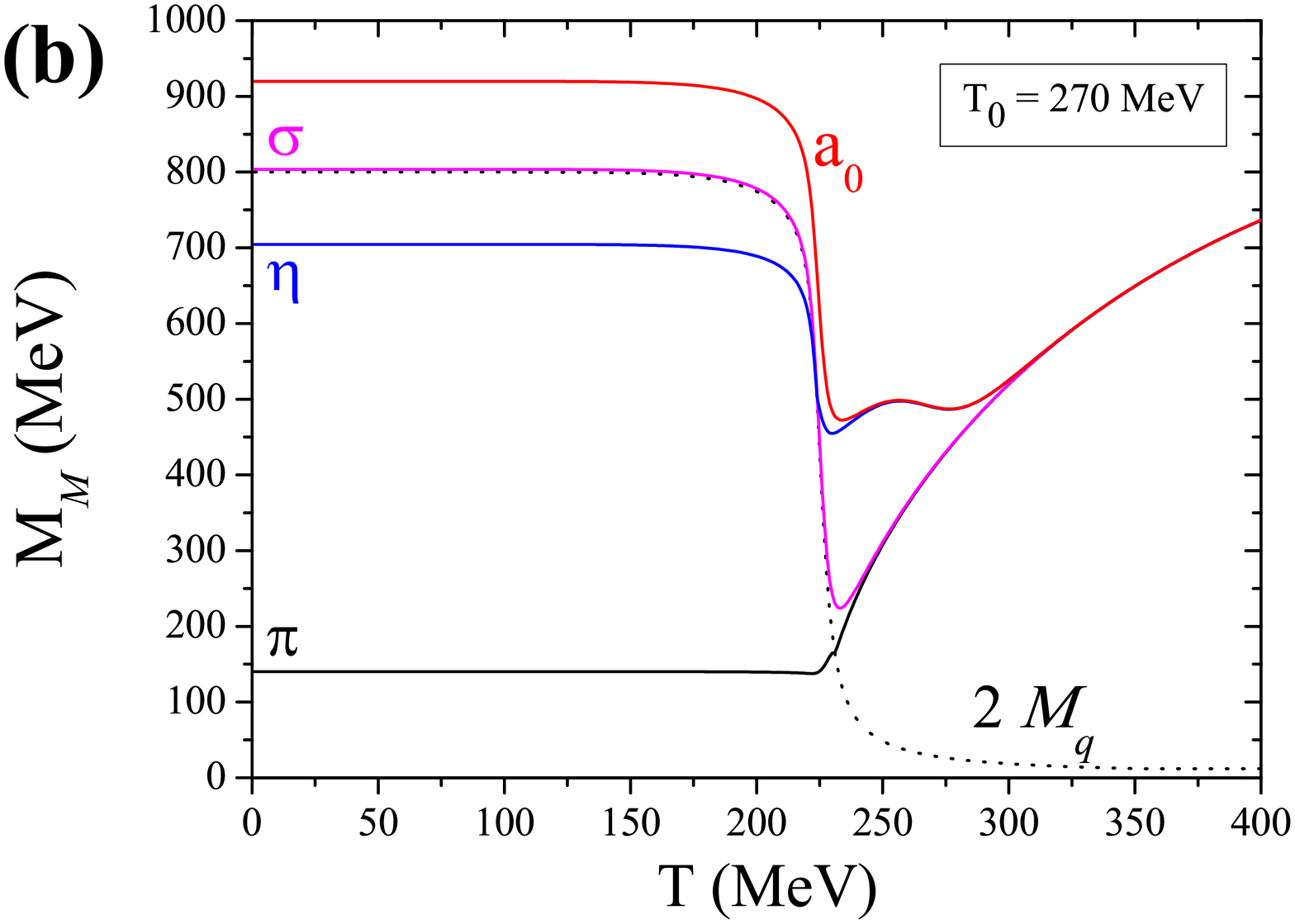,width=5.5cm,height=4.5cm}\\
       \end{tabular}
 \end{center}
\caption {Meson masses in  PNJL (upper panels) and  EPNJL (lower panels) models.}
\label{fig:PNJL}
\end{figure}
%%%%%%%%%%%%%%%%%%%%%%%%%%%%%%%%%%%%

Alternatively, we can  use the redefinition of Eq. (\ref{redef}), with  
$c(T)= 0.2 f(T), \,\,{\rm where} \,\,\, f(T) = 1/\left(1+\exp\left((T -T_0)/10\right)\right)$.
In the EPNJL model we have an equivalent explicit temperature dependence: 
$g_1 (\Phi, T) = G(\Phi) (1 - c(T)), \,\,\,\,\,\,g_2 (\Phi, T)= G(\Phi)\,c(T)$ (scenario B).
Both scenarios insure the vanishing of the anomaly and, therefore, the convergence of axial 
partners, but in  scenario A, for both models,  there is a unphysical region of negative
pressures for low values of $T_0$. This problem that does not exist in scenario B, that allows 
more freedom in the choice of $T_0$. Here we will only discuss this last scenario.

In Fig. \ref{fig:PNJL} we plot the PNJL results  for the meson masses,
considering two values of $T_0$.  For $T_0 = 270$ MeV, the convergence of chiral
partners occurs first than that of axial partners, as usual, with $T_{eff}\approx 300$ MeV. 
The new finding is that for $T_0 = 210$ MeV the  chiral and axial partners get degenerate very 
closely and  $T_{eff}$ is lower.  
Concerning the topological susceptibility, it allways vanishes, this effect being driven by   
the vanishing of the quark condensate.

The behavior of the meson masses and topological susceptibility in the EPNJL model is
qualitatively similar to PNJL model, as  it can be seen in Fig. 1.
Here also the restoration of chiral and axial symmetries become closer for low values of $T_0$,
but the temperatures for the effective restoration of symmetries are slightly lower.
We conclude that in EPNJL there is entanglement between  deconfinement and restoration of chiral symmetry
but not with restoration of axial symmetry. 
Restoration of axial symmetry requires an adittional mechanism of instanton suppression.

%%%%%%%%%%%%%%%%%%%%%%%%%%%%%%%%%%%%%%%%%%%%%%%%%%%%%%%%%%%%
\vskip0.25cm
Work supported by projects CERN/FP/116356/2010 and PTDC/FIS/ 100968/2008, 
 developed under the initiative QREN financed by  UE/FEDER through
the Programme COMPETE - ``Programa Operacional Factores de Competitividade''.

%%%%%%%%%%%%%%%%%%%%%%%%%%%%%%%%%%%%%%%%%%%%%%%%%%%%%%%%%%%%
%%%%%%%%%%%%%%%%%%%%%%%%%%%%%%%%%%%%%%%%%%%%%%%%%%%%%%%%%%%%
%%%%%%%%%%%%%%%%%%%%%%%%%%%%%%%%%%%%%%%%%%%%%%%%%%%%%%%%%%%%

\end{document}